# Optical phonon dynamics and electronic fluctuations in the Dirac semimetal $Cd_3As_2$

A. Sharafeev[1,2], V. Gnezdilov[1,3], R. Sankar[4,5], F. C. Chou[4], and P. Lemmens[1,2*]

[1]*Inst. for Condensed Matter Physics, TU Braunschweig, D-38106 Braunschweig, Germany*

[2]*Laboratory for Emerging Nanometrology, TU Braunschweig, D-38106 Braunschweig, Germany*

[3]*B. I. Verkin Inst. for Low Temperature Physics and Engineering, National Academy of Sciences of Ukraine, Kharkov 61103, Ukraine*

[4]*Center for Condensed Matter Sciences, National Taiwan University, Taipei 10617, Taiwan*

[5]*Institute of Physics, Academia Sinica, Nankang, Taipei 11529, R.O.C. Taiwan*

## Abstract

Raman scattering in the three-dimensional Dirac semimetal $Cd_3As_2$ shows an intricate interplay of electronic and phonon degrees of freedom. We observe resonant phonon scattering due to interband transitions, an anomalous anharmonicity of phonon frequency and intensity, as well as quasielastic (E~0) electronic scattering. The latter two effects are governed by a characteristic temperature scale $T^* \sim 100$ K that is related to mutual fluctuations of lattice and electronic degrees of freedom. A refined analysis shows that this characteristic temperature corresponds to the energy of optical phonons which couple to interband transitions in the Dirac states of $Cd_3As_2$. As electron-phonon coupling in a topological semimetal is primarily related to phonons with finite momenta, the back action on the optical phonons in only observed as anharmonicities via multi-phonon processes involving a broad range of momenta. The resulting energy density fluctuations of the coupled system have previously only been observed in low dimensional or frustrated spin systems with suppressed long range ordering.

# I. INTRODUCTION

Topological semimetals are an interesting and novel state of matter that attracts considerable interest in condensed matter physics. There exist relativistic bulk electronic bands with linear dispersion, non-trivial spin texture due to strong spin-orbit coupling, and other prominent features. One of the representatives of this group of materials is $Cd_3As_2$, which is well known for an anomalously high electron mobility ($\approx 8.0$ m$^2$ V$^{-1}$ s$^{-1}$) despite a semimetallic character of conductivity and a complex, defect-ordered structure[2–5]. There are also other 3D topological Dirac semimetals, such as $A_3$Bi ($A$ = K, Rb, Na)[11] or $\beta$-cristobalite $BiO_2$[12]. However, they are all instable under ambient conditions. In this context, $Cd_3As_2$ stands out with respect to application oriented properties.

Recent studies of $Cd_3As_2$ show two stable Dirac nodes at two special k points along the G-Z momentum space direction with linear dispersing bands in all three dimensions[6,7,8]. In addition, the Fermi velocity of these 3D Dirac fermions is quite high ($\approx 2 \cdot 10^6$ m·s$^{-1}$)[6] in comparison with the well-studied 2D Dirac systems. It is about 3 times higher than in the topological surface states of $Bi_2Se_3$[9] and 1.5 times higher than in graphene[10].

Topological Dirac semimetals can be driven into a Weyl semimetal or a topological insulator by symmetry breaking or increasing spin-orbit coupling, respectively. Reducing dimensionality a quantum spin Hall insulator is induced[13]. Therefore, the presence or absence of a center of inversion in $Cd_3As_2$ is crucial and affects the electronic band structure[14]. The fundamental relationship between the crystalline symmetries, the topological properties of Dirac states and generic phase diagrams are given in Ref. [15]. The existence of a center of inversion in the crystal structure of $Cd_3As_2$ is discussed controversially. This is based on its complex, defect-ordered structure that could even lead to a local symmetry based on preparation conditions of the samples. Therefore, an independent experimental study is important.

Here, we report on results of a Raman scattering study on $Cd_3As_2$ single crystals. This system shows evidence for an intricate interplay of phonons with electronic degrees of freedom leading to a characteristic temperature T*~100 K. This characteristic temperature is related to pronounced fluctuations that are observed in the temperature dependence of optical phonons as well as in quasielastic electronic scattering.

# II. EXPERIMENTAL DETAILS

Single crystals of $Cd_3As_2$ with dimensions of 2 mm$^3$ were grown by self-selecting vapor growth (SSVG) method. Raman scattering measurements were performed in quasi-backscattering geometry from shiny surface of the crystals with a triangular shape which corresponds to the {112} plane[16]. Samples were cleaned with acetone and isopropanol before fixing them on the sample holder using silver glue. Raman spectra were measured in both parallel (*xx*) and crossed (*yx*) polarizations, with incident light polarized perpendicular or parallel to the base of the triangle, respectively.

As excitation sources, solid state ($\lambda$= 532 and 488 nm) and Ar-Kr-ion ($\lambda$= 488, 514.5, 568, and 647 nm) lasers were used. A laser power of less than 10 mW was focused to a 100-$\mu$m-diameter spot on the crystal. Note that in the present experiments we used a power of the exciting radiation that is at least 20 times smaller than in previous Raman experiments on $Cd_3As_2$[17,18].

The spectra were collected via a triple spectrometer (Dilor-XY-500) by a liquid nitrogen cooled CCD (Horiba Jobin Yvon, Spectrum One CCD-3000V). Temperature dependencies from 9 K to 300 K of the Raman spectra were measured in a variable temperature closed cycle cryostat (Oxford/Cryomech Optistat).

### III. RESULTS

*Crystal structure and phonon spectra.* The crystal structure of $Cd_3As_2$ is complicated and can be related to tetragonally distorted antifluorite structure with 1/4 Cd site vacancies. At high temperatures ($T$ > 873 K), the distribution of these vacancies is random and leads to the ideal antifluorite space group $Fm\bar{3}m$. In the temperature region of 873 K > $T$ > 648 K, Cd ions are ordered leading to $P4_2/nmc$[14]. At lower temperatures, the crystal structure of $Cd_3As_2$ was initially determined as non-centrosymmetric ($I4_1cd$)[19] and finally proposed to be centrosymmetric ($I4_1acd$)[14]. Resolving the ambiguity between centrosymmetric and noncentrosymmetric crystal structures can be difficult in some cases and this is a well-known problem in crystallography[20,21]. Raman scattering is an effective tool to refine the crystal structure based on the number or observed phonon modes.

The factor group analysis yields 145 Raman-active phonon modes ($\Gamma_{Raman}$ = 26$A_1$ + 27$B_1$ + 27$B_2$ + 65$E$) for noncentrosymmetric $I4_1cd$ (No 110, Z = 16) and 74 modes for centrosymmetric $I4_1acd$ (No 142, Z=32). In previous a previous Raman study only nine phonon modes have been observed[17,18].

Since our Raman measurements were done on a natural cleavage {112} plane, all possible Raman-active phonon modes should be observable. Using a $\lambda$ = 532 nm laser excitation line with a power of 20 mW there are 13 modes at room temperature and 27 lines at low temperatures, see Fig. 1, i.e. a larger number than in previous studies. Using a $\lambda$ = 647 nm excitation 44 phonon modes are observed. Corresponding spectra and a table are given in the supplement. The observed phonons in $Cd_3As_2$ that can be divided into two frequency regimes, below 80 cm$^{-1}$ and 150 – 250 cm$^{-1}$. We attribute thee regimes to collective vibrations of mainly Cd coordinations and As phonons, respectively. The spectra in (*xx*) polarization and at high temperatures are superposed by a pronounced and extended quasielastic (QE) signal. This scattering is an important evidence for fluctuations and will be discussed further below.

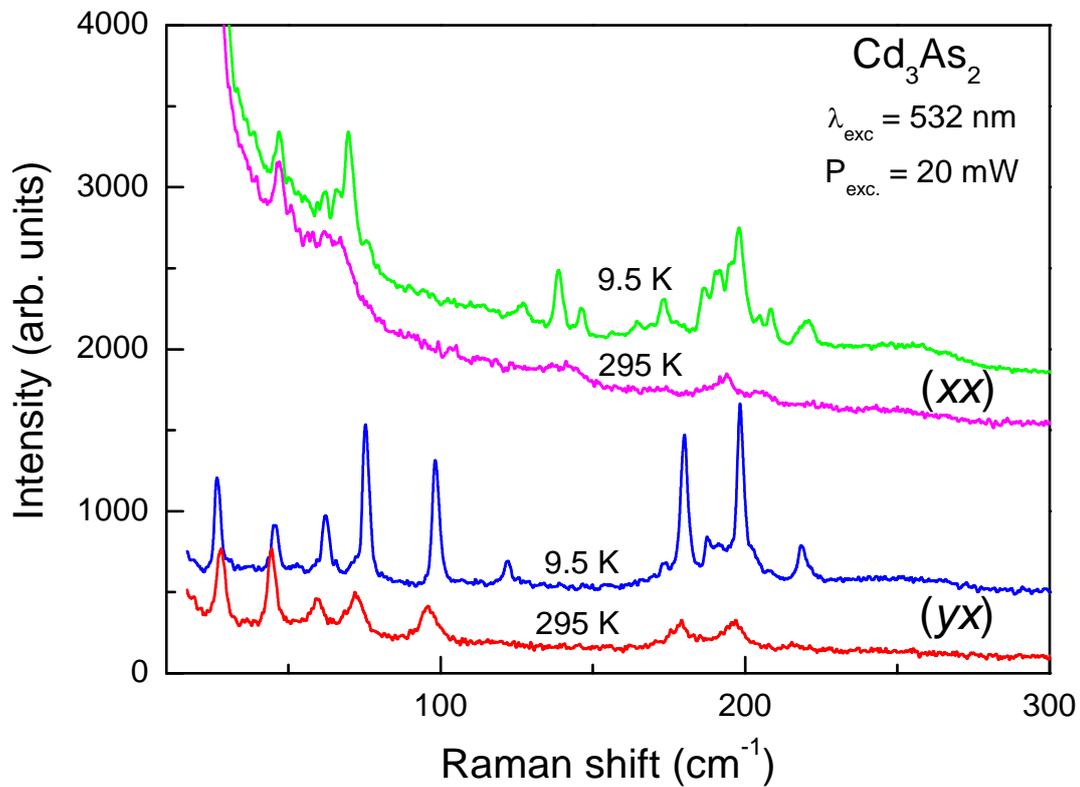

Figure 1. Raman spectra of Cd$_3$As$_2$ taken with λ = 532 nm excitation line, (*xx*) and (*xy*) polarizations, at T= 9.5 K and room temperature, respectively.

Fig. 2 documents a pronounced dependence of the Raman scattering intensity on the energy of the incident radiation. The inset of this figure gives the intensity of the 150-250 cm$^{-1}$ integrated phonon intensity and of the QE scattering intensity. The phonons show the largest intensity at with an excitation at low energies (λ = 647 nm, 1.9 eV). This wavelength coincides with an intrinsic electronic transition of Cd$_3$As$_2$ as observed in reflectivity spectra of Cd$_3$As$_2$[1]. The QE scattering shows the largest intensity at the largest energies (λ = 488 nm, 2.54 eV). This difference indicates that distinctively different electronic states are involved in the Raman intermediate states.

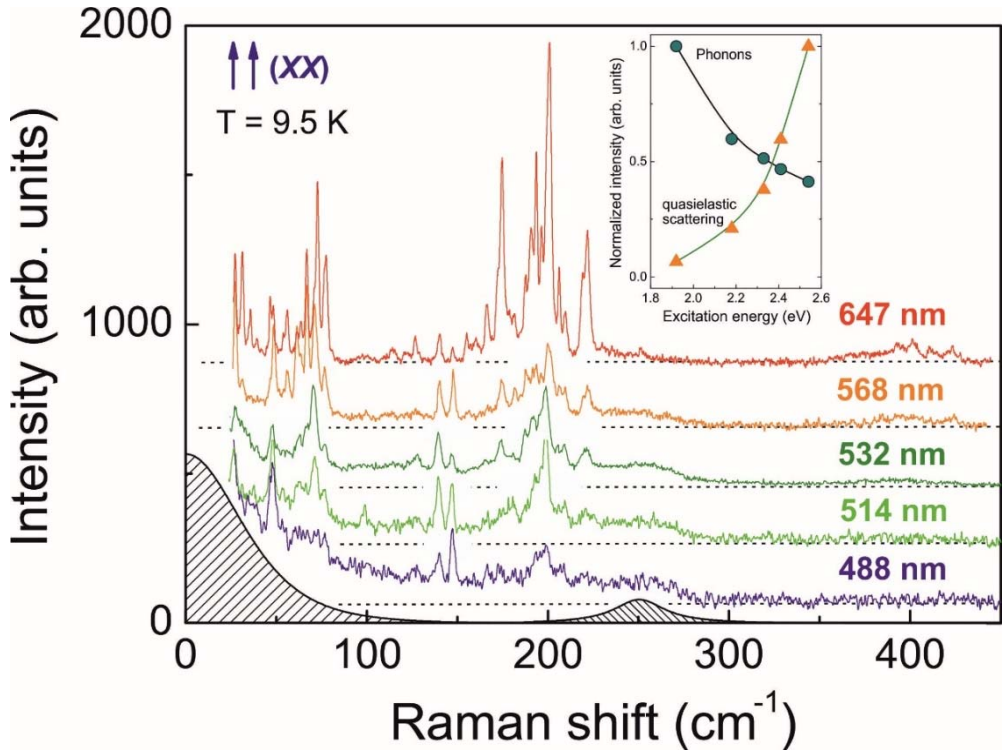

Figure 2. Raman spectra of $Cd_3As_2$ with different incident laser excitations, (*xx*) polarization, and $T = 9.5$ K. The dashed lines correspond to the background. The dashed areas represent a fitting of the QE scattering and a plasmon maximum at 250 cm$^{-1}$ to the 488-nm data, as described in the text. The inset shows the normalized intensity of the QE scattering and the integrated 150-250-cm$^{-1}$ phonon scattering as function of excitation energy.

In the following we will focus on Raman spectra with λ = 647 nm excitation as the phonon lines are well distinguished and have largest intensities. Selected temperature dependent Raman spectra of $Cd_3As_2$ are shown in Fig. 3. A detailed list of the phonon modes is given in the supplement, Table SI. This data is based on fits by Lorentzian profiles given in a figure in the supplement. The number of 44 observed phonon modes allows us to support the centrosymmetric ($I4_1acd$) structure since the non-centrosymmetric ($I4_1cd$) structure should show a significantly larger number of phonon modes in Raman scattering.

In Fig. 3 we demonstrate that there exist pronounced phonon anomalies as function of temperature. For the low energy modes this can be easily quantified by fitting, see Fig. 4. For the higher frequency phonons also anomalous properties are noted, especially a strong broadening of their linewidth. However, due to the close proximity of the modes an individual fitting is not viable. In addition there exist two phonon scattering at approximately 400 cm$^{-1}$.

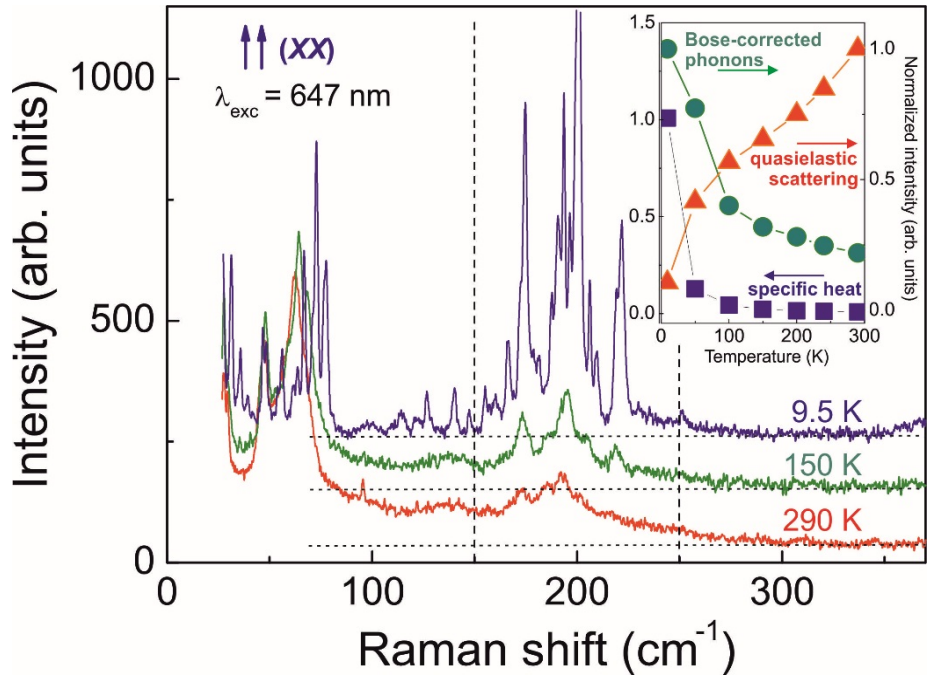

Figure 3. Raman spectra of $Cd_3As_2$ at three selected temperatures measured in (*xx*) polarization and λ = 647-nm excitation. The data is shifted and dashed lines mark the background. The inset shows the temperature dependence of the normalized intensity of the QE scattering, the fluctuation-induced specific heat $C_p^{fluct}$, and the Bose-corrected phonon scattering intensity, respectively. The crossing point of the QE scattering and the phonon intensity marks a characteristic energy scale T* ~ 100 K.

In Fig. 4 the temperature dependence of the phonon frequencies ω(*T*) and linewidths Γ(*T*) (FWHM) of selected phonon modes (46.5, 48.5, 67 and 73 cm$^{-1}$) are given. There exist pronounced changes in slope that are not in accordance with the expected anharmonicity of optical phonons based on multi-phonon scattering processes. The latter is a continuous function resulting from anharmonic interactions by three- and four-phonon scattering and decay processes (dashed lines)[22–24]. It does not show a turning point and can be fitted to the high temperature evolution of frequency and linewidth. The deviations between the modelled anharmonicity and the experimental data are substantial and should be regarded as strong evidence for a relaxation mechanism different from multiphonon states. The resulting deviations are shown in the insets of Fig. 5a and b with a broad maximum and a sharper decrease for temperatures T< T* ~ 100 K. This dependence is obviously not due to some kind of critical behaviour. As the intensity of quasielastic electronic scattering decreases in the same temperature regime we consider a coupling to electronic degrees of freedom as the origin for the anomalous anharmonicity. This will be discussed further below.

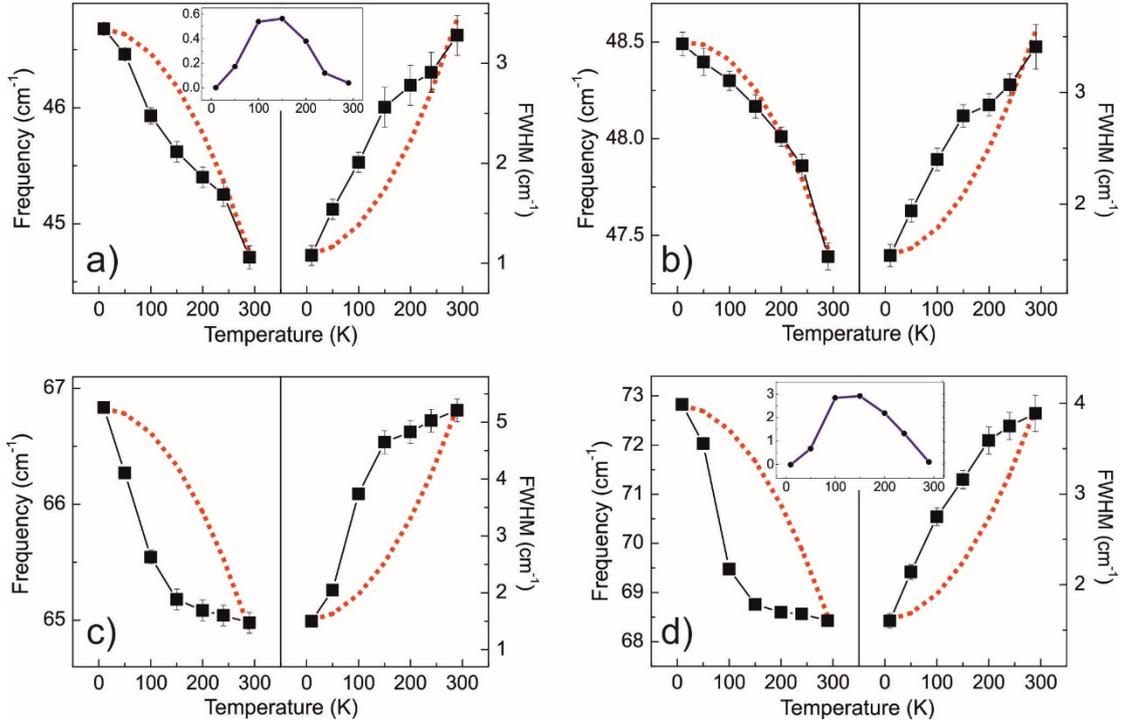

Figure 4. Temperature dependence of the phonon frequencies and linewidths for the modes at (a) 46.5 cm$^{-1}$, (b) 48.5 cm$^{-1}$, (c) 67 cm$^{-1}$, and (d) 73 cm$^{-1}$. The dashed red lines approximate an expected anharmonic variation of phonons with temperature[22–24]. The insets in a) and d) show the respective deviation of measured and approximated phonon frequency with temperature.

*Quasielastic scattering.* In the following we will focus on QE scattering in Cd$_3$As$_2$ with data presented in Fig. 1 to Fig. 3. This contribution dominant in (*xx*) polarization and has a linewidth below approximately 100 cm$^{-1}$. Its integrated spectral weight is largest at high temperatures and shows a strong suppression at low temperatures where the phonon scattering intensity increases, see inset of Fig. 3. At T* ~ 100 K these two counteracting temperature dependencies cross each other. The lineshape of the QE scattering is fitted using the λ=488 nm-data with the following equation, $I(\omega) = I_o / ((((2\sqrt[N]{2} - 1)(\omega - \omega_0)/\gamma)^2 + 1)^N)$, where $I_0$ is the intensity at $\omega_0$, the center of the maximum. With the parameter $N = 1$ a pure Lorentzian line shape is resolved, while large $N$ (~200) leads to a Gaussian line shape. The fit to the experimental data (dashed area in Fig. 2) corresponds to $N \approx 2.7$. Therefore, we conclude a Lorentzian line shape of this quasielastic process.

In semimetals such Lorentzian QE scattering is of electronic origin and can be mapped on energy density fluctuations[28]. In such a case, a related specific heat is given by the scaled intensity of the quasielastic scattering, $C_p^{fluct}(T) \propto I_{quasiel} / T^2$ [25], see inset of Fig. 3. This quantity is strongly increasing at low temperatures. Finally, QE scattering shows a resonant enhancement as function of incident excitation which is different from the phonon scattering[26]. This implies that phonon scattering and QE scattering involve different intermediate states in the Raman scattering process[26,27].

## IV. DISCUSSION

There is a similarity in the observation of QE scattering in $Cd_3As_2$ and observations in low dimensional spin systems[29]. In the latter related fluctuations are due to the suppression of long range magnetic order. Such systems, e.g. $(VO)_2P_2O_7$ [30], $\alpha'$-$NaV_2O_5$ [31], and $SrCu_2(BO_3)_2$ [32] also show an enhanced anharmonicity of the optical phonons. Generally speaking, these anomalies are related to the tendency of the spin or electron system that is coupled to a phonon bath to reduce the overall degeneracies of the system. Following this concept, it makes sense that for temperatures below T* ~ 100K the QE scattering decreases, the phonon scattering intensity increases, and phonon anharmonicity changes to an expected behavior.

At this point it is useful to consider again the specific properties of the phonon system of $Cd_3As_2$. We expect that the phonons with frequencies below approximately 80 cm$^{-1}$ ($\equiv$ 110 K), which fits in energy to T*, couple to Dirac states close to the Fermi energy. This is due to their relation to Cd ions and vacancies which lift or preserve the inversion symmetry needed for the Dirac semimetal. In contrast the phonons at 150-250 cm$^{-1}$ ($\equiv$ 220-360K), above the quasi-gap of phonon excitations, correspond to As vibrations not expected to be relevant for the Dirac states. Considering a thermal population of the low energy phonons, a temperature variation in the range of the quasi-gap of phonons will lead to strong variations of the occupied states. Thereby all physical properties related to the phonon density of states will be affected. These are transport properties, like resistivity, but also the anharmonicity of optical phonon. This effect is further amplified due to some randomness of the Cd vacancy sites.

On the other hand theory proposes a back action of the phonon self-energy to electronic states going beyond the above mentioned phonon bath[33]. It leads to a modification of electron-phonon matrix elements with band inversion[34]. In a topological insulator intraband (particle-hole pair) scattering will induce an enhanced phonon linewidth and variations of the phonon frequency for phonons close to the Brillouin zone center. Such topologically-induced phonon shifts are of the order of a few cm$^{-1}$ [33], in good agreement with experiment[35–37]. For a topological semimetal interband scattering and phonons with larger momenta are important. To our knowledge there exist no direct experimental study of topological effects on the phonon dispersion at finite k. However, the anharmonicity of optical phonons involves a scattering processes into multiphonon states with a broad range of momenta. In this way interband scattering processes in electronic states with band inversion can contribute to the self-energy of optical phonons. To allow such a coupling energy scales of the respective states have to match and the phonon states have to be populated. Therefore our studies show largest anharmonicities as well as intensity gains for phonons which are very close to the lower and upper threshold of the quasi-gap and the characteristic temperature, $T^* \sim 100$ K, matches to the frequency of these phonons. We propose that this degeneracy allows enhanced fluctuations.

In transport studies on $Cd_3As_2$ microbelts with ultrahigh mobility similar characteristic temperature scales have been demonstrated. Using a Drude-like approach the derived carrier density shows a minimum at 150K and a following increase towards low temperatures[38]. In another experiment an anomalously large Nernst effect is only observed below 50 K and attributed to the temperature evolution of the transport relaxation rate[39]. In Hall effect measurements on nanoplates the largest changes with field are observed between 80 and 100 K together with a transition in the type of conduction[40]. In a critical discussion of these effects, however, the effect of doping, i.e. the shift of the $E_F$ from the Dirac cone should be taken into account. In a recent investigation of optical conductivity it has been shown that a charge carrier crossover at $T^* = 100$ K is compatible with a Fermi energy of $E_F \approx 25$ meV[41] due to thermally excited carriers.

Finally, we address direct evidence for scattering processes on the Dirac states using the Raman process, as single particle or collective excitations (plasmon-like). Such processes evade screening either by an electronic mass anisotropy or strong defect scattering. Evidences for electronic Raman scattering in Dirac electron systems have previously been found in carbon nanotubes[42] and Rashba semimetals[43]. The corresponding effects consist of broad Gaussian maxima and range in energy from $\Delta\omega = 40$ cm$^{-1}$ to 500 cm$^{-1}$. In $Cd_3As_2$ we observe one Gaussian maximum at $\omega_{plas} = 250$ cm$^{-1}$ with a width (FWHM) of $\Gamma = 30$ cm$^{-1}$ that fits to this scenario. In the optical conductivity Dirac states contribute in a very different way. Here a linear additional component is expected with an onset for frequencies $\omega > \omega_{Pauli} = 2E_F$, which is the Pauli-blockade energy. In very thin crystals of the Dirac semimetal $ZrTe_5$ such an effect has been observed at 120 cm$^{-1}$ [44]. In $Cd_3As_2$ such an onset has not been resolved up to now due to a strong Drude contribution[45].

## IV. CONCLUSIONS

In our Raman scattering study of the three-dimensional Dirac semimetal $Cd_3As_2$ with centrosymmetric structure (*I*4$_1$*acd*) a resonant enhancement of phonon scattering is observed with an excitation of 1.9 eV, while quasielastic electronic scattering shows a different energy dependence. This energy corresponds to the energy of the interband electronic transition observed earlier in reflectivity measurements. The temperature dependence of phonon and quasielastic scattering counteract defining a characteristic temperature scale T* ~ 100 K with anomalous anharmonicities of the phonon frequency and linewidth. We attribute these effects to interband scattering at the Dirac states coupling to phonon anharmonicity and pronounced energy density fluctuations of the coupled electron-phonon system. T* ~ 100 K corresponds to a matching of phonon with thermal energies.

## ACKNOWLEDGEMENTS

We thank David Schmeltzer and Ion Garate for important discussions. We would like to acknowledge the support of the International Graduate School of Metrology (B-IGSM) and the Graduate School Contacs in Nanosystems. Financial support by GIF and DFG-RTG 1952 Nanomet, and DFG-LE967/16-1 is appreciated. FCC acknowledges support from the Ministry of Science and Technology in Taiwan under grant No. MOST-104-2119-M-002 -028-MY2.

Supplement to

# Optical phonon dynamics and electronic fluctuations in the Dirac semimetal Cd$_3$As$_2$

A. Sharafeev, V. Gnezdilov, R. Sankar, F. C. Chou, and P. Lemmens

In this supplement details of phonon Raman scattering supporting a centrosymmetric (I41acd) crystal structure of Cd3As2 are documented. Fig. S1 shows Raman spectra at T=9.5K with 647-nm excitation including fits using Lorentzian lines. A number of 44 modes are identified and listed in Tab. S1. The non-centrosymmetric (I41cd) structure should show a significantly larger number of phonon modes.

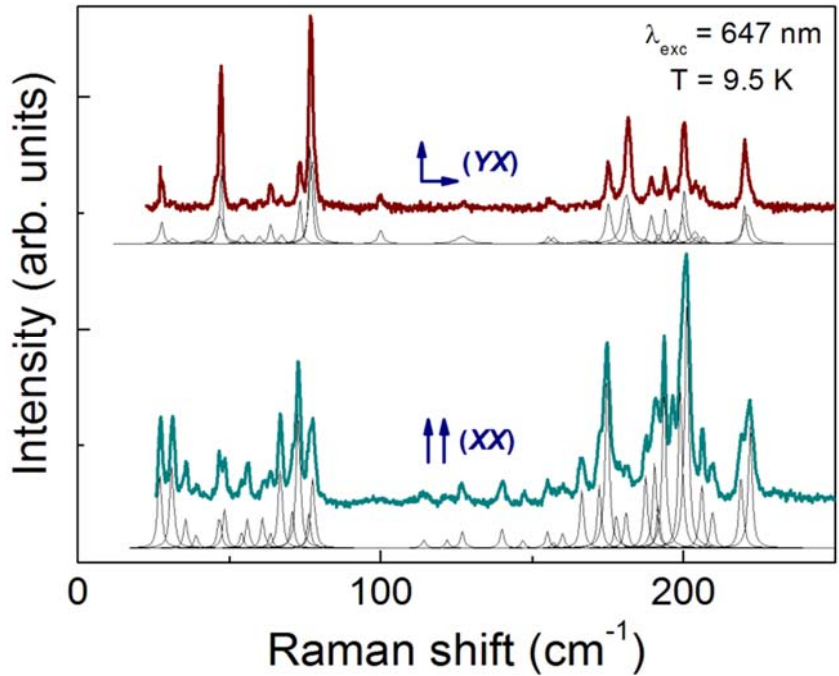

Figure S1. Low temperature Raman spectra of Cd$_3$As$_2$ at T = 9.5 K in (xx) and (xy) polarizations. Black lines show fits to the phonon modes from Table S1 using Lorentzians.

To describe the anomalous anharmonicities we refer to a model that considers the decay of optical phonons due to cubic anharmonic interactions[22–24]. Within this approximation occupation factors of frequency averaged, decaying phonons $\omega_{av3}$ and $\omega_{av4}$ are weighted by scaling factors $b_j$ and $c_j$, and used to describe three- and four-particle processes:

$$\omega_j(T) = \omega_{0j} - \frac{b_j}{\exp(\hbar\omega_{av3}/k_BT)-1} - \frac{c_j}{\exp(\hbar\omega_{av4}/k_BT)-1} - \frac{c_j}{\left[\exp(\hbar\omega_{av4}/k_BT)-1\right]^2},$$

here, $\omega_{0j}$ is the eigenfrequency of the phonon at $T = 0$ K. In simplest approximation the frequency of the decaying phonons are $\omega_{av3} \approx \omega_{0j}/2$ and $\omega_{av4} \approx \omega_{0j}/3$).

The phonon linewidth is described in a related way,

$$\Gamma_j(T) = \Gamma_{0j} + \frac{d_j}{\exp(\hbar\omega_{0j}/2k_BT)-1} + \frac{f_j}{\exp(\hbar\omega_{0j}/3k_BT)-1} + \frac{f_j}{\left[\exp(\hbar\omega_{0j}/3k_BT)-1\right]^2},$$

where $\Gamma_0$, $\omega_0$, $d_j$, and $f_j$ are the width, eigenfrequency at $T = 0$ K, and mode dependent parameters for three- and four-particle decay and coalescence processes, respectively.

Table I. List of the 44 phonon modes of $Cd_3As_2$ determined at T=9.5K and $\lambda$ = 647 nm excitation. Highest intensity is denoted by "+++", while "-" denotes zero intensity.

| w (cm⁻¹) | XX | XY | Assignment | w (cm⁻¹) | XX | XY | Assignment |
|---|---|---|---|---|---|---|---|
| 27.4 | +++ | + | $B_{1g}$ | 155 | ++ | + | $B_{2g}$ |
| 31.2 | +++ | - | $A_{1g}$ | 157.2 | + | + | $B_{2g}$ |
| 35.6 | + | - | $A_{1g}$ | 160.1 | + | - | $A_{1g}$ |
| 39.3 | ++ | + | $B_{2g}$ | 166.4 | +++ | + | $B_{1g}$ |
| 46.7 | + | ++ | $E_g$ | 172.2 | + | - | $A_{1g}$ |
| 48.3 | + | + | $B_{2g}$ | 174.7 | +++ | + | $B_{1g}$ |
| 53.9 | ++ | + | $B_{2g}$ | 177.8 | + | - | $A_{1g}$ |
| 56 | +++ | + | $B_{1g}$ | 181.4 | + | ++ | $E_g$ |
| 59.9 | - | + | $E_g$ | 187.5 | +++ | + | $B_{1g}$ |
| 61.5 | + | - | $A_{1g}$ | 190.4 | +++ | + | $B_{1g}$ |
| 63.6 | + | ++ | $E_g$ | 191.6 | ++ | + | $B_{2g}$ |
| 66.8 | +++ | + | $B_{1g}$ | 193.6 | +++ | + | $B_{1g}$ |
| 70.8 | + | - | $A_{1g}$ | 196.4 | +++ | + | $B_{1g}$ |
| 72.8 | +++ | + | $B_{1g}$ | 199.2 | +++ | + | $B_{1g}$ |
| 76.3 | + | ++ | $E_g$ | 201 | +++ | + | $B_{1g}$ |
| 77.6 | + | ++ | $E_g$ | 204.1 | - | + | $E_g$ |
| 98.7 | + | ++ | $E_g$ | 206.3 | +++ | + | $B_{1g}$ |
| 114.2 | + | - | $A_{1g}$ | 209.6 | + | - | $A_{1g}$ |
| 121.7 | + | - | $A_{1g}$ | 218.9 | + | - | $A_{1g}$ |
| 126.8 | + | + | $B_{2g}$ | 220.2 | - | + | $E_g$ |
| 140 | + | - | $A_{1g}$ | 220.8 | - | + | $E_g$ |
| 147.4 | + | - | $A_{1g}$ | 221.8 | + | - | $A_{1g}$ |